\documentclass[onecolumn]{revtex4}
\usepackage{ragged2e}
\usepackage{amsmath}
\usepackage{graphicx}
\usepackage{bm}

\usepackage[usenames,dvipsnames]{color}

\definecolor{darkblue}{RGB}{0,0,196}

\begin{document}

\title{Comparing effective temperatures in standard, Tsallis, and q-dual
statistics from transverse momentum spectra of identified light
charged hadrons produced in gold--gold collisions at RHIC
energies}\vspace{0.5cm}

\author{Ting-Ting~Duan$^{1,}$\footnote{\, 202312602001@email.sxu.edu.cn},
Pei-Pin~Yang$^{2,}$\footnote{\, peipinyangshanxi@163.com;
peipinyang@xztu.edu.cn}, Peng-Cheng~Zhang$^{1,}$\footnote{\,
202312602003@email.sxu.edu.cn}, Hai-Ling~Lao$^{3,}$\footnote{\,
hailinglao@163.com; hailinglao@pku.edu.cn},
Fu-Hu~Liu$^{1,}$\footnote{\, Correspondence: fuhuliu@163.com;
fuhuliu@sxu.edu.cn}, Khusniddin~K.~Olimov$^{4,5,}$\footnote{\,
Correspondence: khkolimov@gmail.com; kh.olimov@uzsci.net}}

\affiliation{$^1$Institute of Theoretical Physics, State Key
Laboratory of Quantum Optics and Quantum Optics Devices \& \\
Collaborative Innovation Center of Extreme Optics, Shanxi
University, Taiyuan 030006, China
\\
$^2$Department of Physics, Xinzhou Normal University, Xinzhou
034000, China
\\
$^3$Department of Science Teaching, Beijing Vocational College of
Agriculture, Beijing 102442, China
\\
$^4$Laboratory of High Energy Physics, Physical-Technical
Institute of Uzbekistan Academy of Sciences, Chingiz Aytmatov Str.
2b, Tashkent 100084, Uzbekistan
\\
$^5$Department of Natural Sciences, National University of Science
and Technology MISIS (NUST MISIS), Almalyk Branch, Almalyk 110105,
Uzbekistan}

\begin{abstract}

\vspace{0.5cm}

\noindent {\bf Abstract:} This study investigates the transverse
momentum ($p_T$) spectra of identified light charged hadrons
produced in gold--gold (Au+Au) collisions across various
centrality classes at center-of-mass energies per nucleon pair,
$\sqrt{s_{NN}}$, ranging from 7.7 to 200 GeV, as measured by the
STAR Collaboration at the Relativistic Heavy Ion Collider (RHIC).
The analysis employs standard (Bose-Einstein/Fermi-Dirac),
Tsallis, and q-dual statistics to fit the same $p_T$ spectra and
derive distinct effective temperatures: $T_{\text{Standard}}$,
$T_{\text{Tsallis}}$, and $T_{\text{q-dual}}$. In most instances,
there exists an approximately linear relationship or positive
correlation between $T_{\text{Tsallis}}$ and
$T_{\text{Standard}}$, as well as between $T_{\text{q-dual}}$ and
$T_{\text{Standard}}$, when considering $T_{\text{Standard}}$ as a
baseline. However, while both $T_{\text{Tsallis}}$ and
$T_{\text{q-dual}}$ increase from semi-central to central Au+Au
collisions at 62.4 GeV and 200 GeV, where QGP is expected, changes
in $T_{\text{Standard}}$ occur more gradually. This work suggests
that $T_{\text{Standard}}$ is better suited for characterizing
phase transitions between hadronic matter and QGP compared to
$T_{\text{Tsallis}}$ or $T_{\text{q-dual}}$, primarily due to the
considerations related to entropy index in the Tsallis and q-dual
statistics.
\\
\\
\end{abstract}

\maketitle

\parindent=15pt

\section{Introduction}

As one of the most fundamental concepts, temperature is
extensively utilized in both scientific research and everyday
life~\cite{1}. To obtain temperature values, a thermometer is
essential. Different types of thermometers exist based on various
temperature scales, allowing for conversion between the
measurements obtained from different devices. However, not all
thermometers are capable of measuring temperatures across a broad
range; indeed, some have very limited measurement capabilities. To
address this limitation and measure temperatures over an extensive
range, multiple thermometers can be employed concurrently. This
necessitates coordination and conversion among the readings from
different thermometers to facilitate meaningful comparisons.

In particular, applying the concept of temperature in high-energy
collisions presents challenges for researchers due to inconsistent
definitions and methodologies. Various types of temperatures exist
within this context, including but not limited to initial state
temperature, chemical freeze-out temperature, kinetic freeze-out
temperature, and effective temperature in high-energy
collisions~\cite{2,3,4,5,6,7}. As their names suggest: initial
state temperature describes the excitation degrees at the onset of
collisions; chemical freeze-out temperature pertains to conditions
when particle interactions cease; while kinetic freeze-out
temperature reflects states where particles no longer interact
significantly during their motion. The effective temperature
indicates the average kinetic energy or transverse momentum of
multiple particles at the stage of kinetic freeze-out while
accounting for flow effects.

Indeed, average kinetic energy arises from both thermal motion
among numerous particles and collective motion originating from
emission sources or groups of particles. The former is
characterized by kinetic freeze-out temperatures whereas the
latter is represented through transverse flow velocity. Here, the
longitudinal motion of the emission source is not considered;
rather, it is assumed that the emission source remains in its rest
frame. Due to limitations in time and space scales, direct
measurements of temperatures in high-energy collisions using
conventional thermometers are not feasible. Instead, the community
employs indirect methods based on yield ratios and transverse
momentum ($p_T$) spectra of multiple particles.

The chemical freeze-out temperature can be ``measured" through
yield ratios. The ``measurements" pertaining to initial state,
kinetic freeze-out, and effective temperatures rely on $p_T$
spectra. Various functions incorporating temperature parameters
can be utilized to fit these $p_T$ spectra; here, the temperature
parameters represent effective temperatures. By introducing
transverse flow velocity into these functions, one can derive the
kinetic freeze-out temperature from the $p_T$ spectra. Utilizing
average $p_T$ ($\langle p_T\rangle$) and root-mean-square $p_T$
($\sqrt{\langle p_T^2\rangle}$), both kinetic freeze-out and
initial state temperatures can be extracted according to
thermal-related methodologies~\cite{8} as well as string
percolation models~\cite{9,10,11}, respectively.

It is important to note that the type of temperature discussed in
high-energy collisions may exhibit model
dependence~\cite{16,17,18,19}. For instance, effective
temperatures derived from standard (Bose-Einstein/Fermi-Dirac),
Tsallis statistics, and q-dual statistics differ
significantly~\cite{20,21,22,23}. While it is acknowledged that
certain thermal or Tsallis distributions can effectively fit
experimental data, exploring the relationships between different
effective temperatures warrants further investigation.
Establishing such relationships will facilitate baseline
comparisons across various models. The primary focus of this work
lies in comparing different effective temperatures.

One can reasonably assume that there exist linear relationships
between \(T_{\text{Tsallis}}\) and \(T_{\text{Standard}}\), as
well as between \(T_{\text{q-dual}}\) and \(T_{\text{Standard}}\),
when considering \(T_{\text{Standard}}\) as a baseline. To
investigate the relationships among different effective
temperatures, we will analyze the production of identified light
charged hadrons (\(\pi^+\), \(\pi^-\), \(K^+\), \(K^-\), \(p\),
and \(\bar p\)) produced in gold--gold (Au+Au) collisions at
center-of-mass energies per nucleon pair of \(\sqrt{s_{NN}} =
7.7\), 11.5, 19.6, 27, 39, 62.4, and 200 GeV. These measurements
were conducted by the STAR Collaboration~\cite{23a,23b,23c} at the
Relativistic Heavy Ion Collider (RHIC). The functions related to
\(p_T\) derived from standard statistics, Tsallis statistics, and
q-dual statistics are employed to extract various effective
temperatures.

The remainder of this paper is organized as follows: Section 2
describes the multi-source picture and formalism associated with
different functions; Section 3 presents our results and
discussion; finally, Section 4 provides a summary and conclusions.

\section{Multi-source picture and formulism}

Within the framework of a multi-source thermal
model~\cite{24,25,26}, it can be posited that multiple emission
sources arise in high-energy collisions due to varying excitation
degrees or interaction mechanisms. This multi-source nature allows
for representation in terms of a multi-component form. For each
individual emission source, distinct models or functions may be
utilized to describe particle production effectively. Naturally
and conveniently chosen for this purpose is the relativistic ideal
gas model within standard statistics for characterizing multiple
particles produced from each source.

The application of the relativistic ideal gas model with
temperature \(T\) yields an expression for the total number of
particles given by~\cite{23}.
\begin{align}
N =gV\int \frac{d^3p}{(2\pi)^3}
\bigg[\exp\bigg(\frac{E-\mu}{T}\bigg)+S\bigg]^{-1},
\end{align}
where $g$ is the degeneracy factor, which equals 1 for bosons such
as $\pi^{\pm}$ and $K^{\pm}$, and 2 for fermions such as $p(\bar
p)$. $V$ represents the volume of the collision system, $p$
denotes momentum,
\begin{align}
E=\sqrt{p^2+m_0^2}=m_T\cosh y
\end{align}
signifies energy, $m_0$ indicates rest mass,
\begin{align}
m_T=\sqrt{p_T^2+m_0^2}
\end{align}
refers to transverse mass,
\begin{align}
y=\frac{1}{2}\ln\bigg(\frac{E+p_z}{E-p_z}\bigg)
\end{align}
stands for rapidity, while $p_z$ corresponds to longitudinal
momentum. The chemical potential of the considered particles is
denoted by $\mu$. In the latter part of this equation, we have
that $S=-1$, which corresponds to Bose-Einstein statistics
applicable to bosons; conversely, when $S=1$, it pertains to
Fermi-Dirac statistics relevant for fermions. As an approximation,
setting $S=0$ aligns with Maxwell-Boltzmann statistics.

The invariant yield or momentum spectrum of the considered
particles is written as~\cite{23}
\begin{align}
E\frac{d^3N}{d^3p}=\frac{1}{2\pi p_T}\frac{d^2N}{dydp_T}
=\frac{1}{2\pi m_T}\frac{d^2N}{dydm_T} =\frac{gV}{(2\pi)^3}E
\bigg[\exp\bigg(\frac{E-\mu}{T}\bigg)+S\bigg]^{-1}.
\end{align}
The density function of momenta can be given by
\begin{align}
\frac{dN}{dp}=\frac{2gV}{(2\pi)^2} p^2
\bigg[\exp\bigg(\frac{\sqrt{p^2+m_0^2}-\mu}{T}\bigg)+S\bigg]^{-1}.
\end{align}
The unit-density function of $y$ and $p_T$ is written as~\cite{23}
\begin{align}
\frac{d^2N}{dydp_T} =\frac{gV}{(2\pi)^2}p_T \sqrt{p_T^2+m_0^2}
\cosh y \bigg[\exp\bigg(\frac{\sqrt{p_T^2+m_0^2} \cosh
y-\mu}{T}\bigg)+S\bigg]^{-1}.
\end{align}

From the unit-density function Eq. (7), the density function of
$p_T$ is given by
\begin{align}
\frac{dN}{dp_T} =\frac{gV}{(2\pi)^2} p_T \sqrt{p_T^2+m_0^2}
\int_{y_{\min}}^{y_{\max}} \cosh y
\bigg[\exp\bigg(\frac{\sqrt{p_T^2+m_0^2}\cosh
y-\mu}{T}\bigg)+S\bigg]^{-1}dy.
\end{align}
Here, $E$ appears in terms of $y$ and $p_T$ due to the integration
for $y$. The experimental rapidity bin $[y_{\min},y_{\max}]$ is
considered as the lower and upper limits of the integration. The
density function of $y$ is
\begin{align}
\frac{dN}{dy} =\frac{gV}{(2\pi)^2} \int_0^{p_{T\max}} p_T
\sqrt{p_T^2+m_0^2}\cosh y
\bigg[\exp\bigg(\frac{\sqrt{p_T^2+m_0^2}\cosh
y-\mu}{T}\bigg)+S\bigg]^{-1}dp_T,
\end{align}
where $p_{T\max}$ denotes the maximum $p_T$ in experiments, though
it is infinity in mathematics.

Considering the multi-source thermal model~\cite{24,25,26} which
results in a multi-component form which reflects a multi-region
structure of $p_T$ spectra~\cite{29,30,31}, one has some
distributions in the multi-component form as followings
\begin{align}
N=\sum_{i=1}^{n_0} N_i,
\end{align}
\begin{align}
E\frac{d^3N}{d^3p}=\sum_{i=1}^{n_0} E\frac{d^3N_i}{d^3p},
\end{align}
\begin{align}
\frac{dN}{dp}=\sum_{i=1}^{n_0} \frac{dN_i}{dp},
\end{align}
\begin{align}
\frac{d^2N}{dydp_T}=\sum_{i=1}^{n_0} \frac{d^2N_i}{dydp_T},
\end{align}
\begin{align}
\frac{dN}{dp_T}=\sum_{i=1}^{n_0} \frac{dN_i}{dp_T},
\end{align}
\begin{align}
\frac{dN}{dy}=\sum_{i=1}^{n_0} \frac{dN_i}{dy},
\end{align}
where $i$ denotes the $i$-th component in $n_0$ components and
$N_i$ is the number of particles in the $i$-th component.

In a multi-component framework, various components can be
expressed in a similar form due to their
commonality~\cite{32,33,34,35} and universality~\cite{36,37,38,39}
observed in high-energy collisions. From the first component
through to the last one within this multi-component structure
reveals a decrease in fraction alongside an increase in
temperature. The average temperature is represented by
\begin{align}
T=\sum_{i=1}^{n_0}k_iT_i,
\end{align}
where $k_i=N_i/N$ is defined as the contribution ratio or fraction
of the $i$-th component among a total of $n_0$ components with
corresponding temperature denoted by $T_i$. Naturally,
$\sum_{i=1}^{n_0}k_i=1$ must hold true.

The multi-component distributions in the standard statistics can
be empirically covered by the distributions in the Tsallis
statistics which has the following related
equations~\cite{18,19,20,21,22}
\begin{align}
N=gV\int \frac{d^3p}{(2\pi)^3}
\bigg\{\bigg[1+\big(q-1\big)\frac{E-\mu}{T}\bigg]^{\frac{1}{q-1}}+S\bigg\}^{-q},
\end{align}
\begin{align}
E\frac{d^3N}{d^3p} =\frac{gV}{(2\pi)^3}E
\bigg\{\bigg[1+\big(q-1\big)\frac{E-\mu}{T}\bigg]^{\frac{1}{q-1}}+S\bigg\}^{-q},
\end{align}
\begin{align}
\frac{dN}{dp}=\frac{2gV}{(2\pi)^2} p^2
\bigg\{\bigg[1+\big(q-1\big)\frac{\sqrt{p^2+m_0^2}-\mu}{T}\bigg]^{\frac{1}{q-1}}+S\bigg\}^{-q},
\end{align}
\begin{align}
\frac{d^2N}{dydp_T} =\frac{gV}{(2\pi)^2}p_T \sqrt{p_T^2+m_0^2}
\cosh y \bigg\{\bigg[1+\big(q-1\big)\frac{\sqrt{p_T^2+m_0^2} \cosh
y -\mu}{T} \bigg]^{\frac{1}{q-1}}+S\bigg\}^{-q},
\end{align}
\begin{align}
\frac{dN}{dp_T} =\frac{gV}{(2\pi)^2} p_T \sqrt{p_T^2+m_0^2}
\int_{y_{\min}}^{y_{\max}} \cosh y
\bigg\{\bigg[1+\big(q-1\big)\frac{\sqrt{p_T^2+m_0^2}\cosh
y-\mu}{T} \bigg]^{\frac{1}{q-1}}+S\bigg\}^{-q}dy,
\end{align}
\begin{align}
\frac{dN}{dy} =\frac{gV}{(2\pi)^2}\int_{0}^{p_{T\max}} p_T
\sqrt{p_T^2+m_0^2} \cosh y
\bigg\{\bigg[1+\big(q-1\big)\frac{\sqrt{p_T^2+m_0^2}\cosh
y-\mu}{T} \bigg]^{\frac{1}{q-1}}+S\bigg\}^{-q}dp_T,
\end{align}
where $q$ is the entropy index which describes the degree of
non-equilibrium of the collision system.

\begin{figure*}[htb!]
\begin{center}
\includegraphics[width=12.0cm]{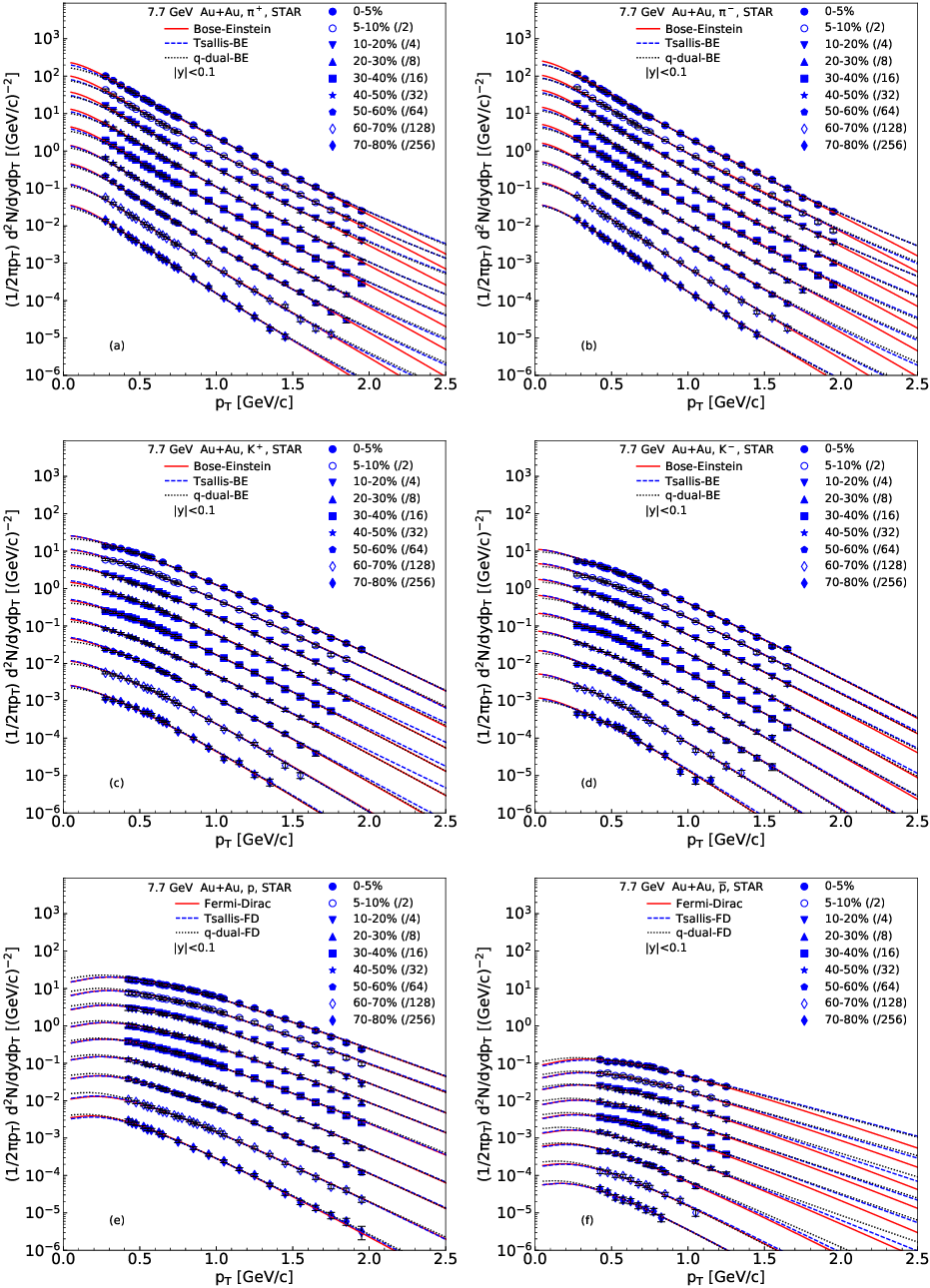}
\end{center}
\justifying\noindent {Figure 1. The invariant yields, $(1/2\pi
p_T)d^2N/dydp_T$, of (a) $\pi^+$, (b) $\pi^-$, (c) $K^+$, (d)
$K^-$, (e) $p$, and (f) $\bar p$ produced in $|y|<0.1$ in Au+Au
collisions at $\sqrt{s_{NN}}=7.7$ GeV. Different symbols represent
the experimental data measured by the STAR
Collaboration~\cite{23a} at RHIC. There are nine centrality
percentage classes (0--5\%, 5--10\%, 10--20\%, 20--30\%, 30--40\%,
40--50\%, 50--60\%, 60--70\%, and 70--80\%) being used. For each
centrality percentage class, the data are scaled by a given
quantity shown in the panel for clear display. The solid, dashed,
and dotted curves are our results fitted by the distributions from
the standard, Tsallis, and q-dual statistics, respectively.}
\end{figure*}

\begin{figure*}[htb!]
\begin{center}
\includegraphics[width=12.0cm]{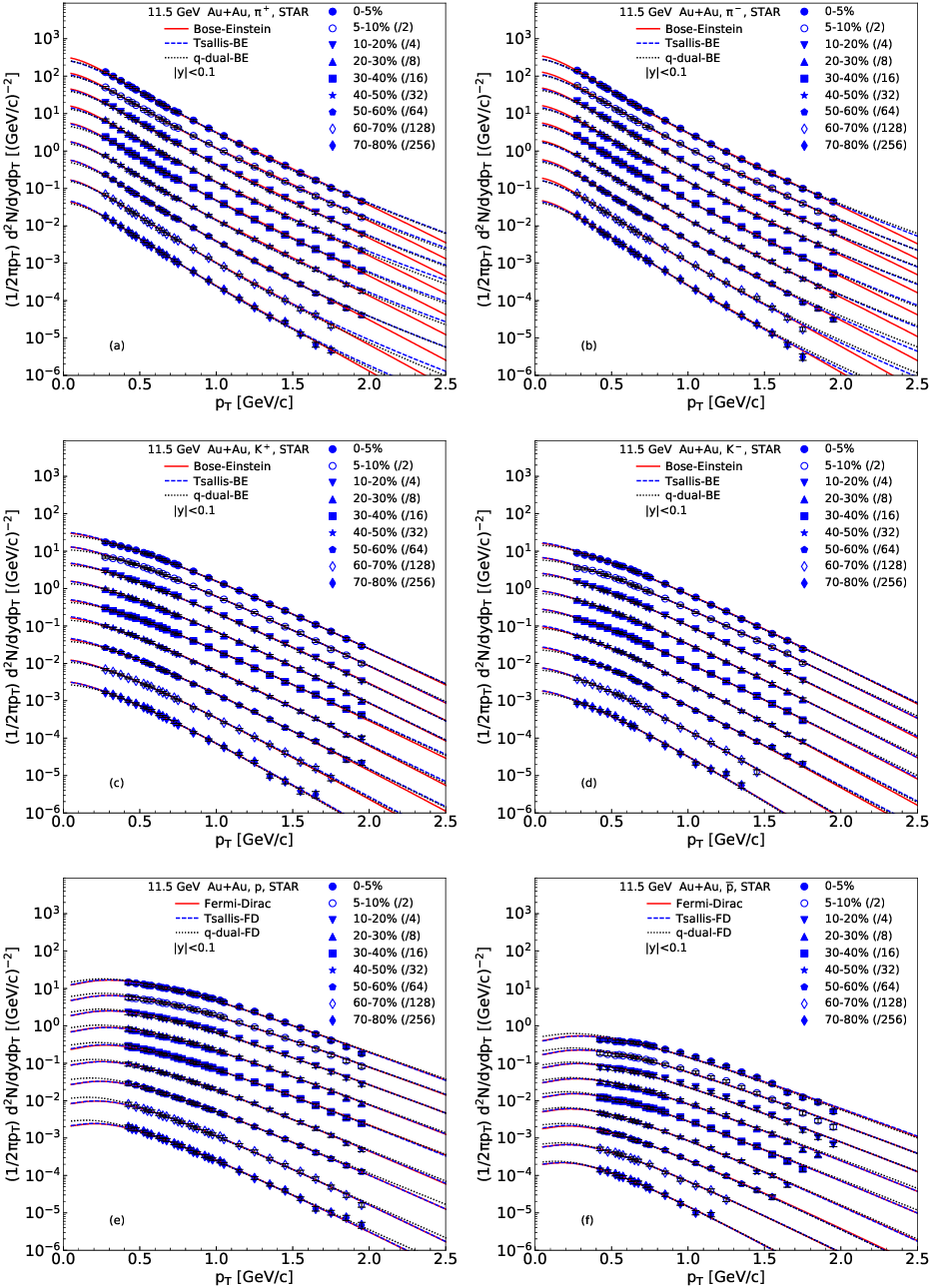}
\end{center}
\justifying\noindent {Figure 2. The invariant yields of (a)
$\pi^+$, (b) $\pi^-$, (c) $K^+$, (d) $K^-$, (e) $p$, and (f) $\bar
p$ produced in $|y|<0.1$ in Au+Au collisions at
$\sqrt{s_{NN}}=11.5$ GeV. Different symbols represent the
experimental data measured by the STAR Collaboration~\cite{23a} at
RHIC with nine centrality percentage classes as those for Figure 1
and scaled by different quantities shown in the panels. The solid,
dashed, and dotted curves are our results fitted by the
distributions from the standard, Tsallis, and q-dual statistics,
respectively.}
\end{figure*}

\begin{figure*}[htb!]
\begin{center}
\includegraphics[width=12.0cm]{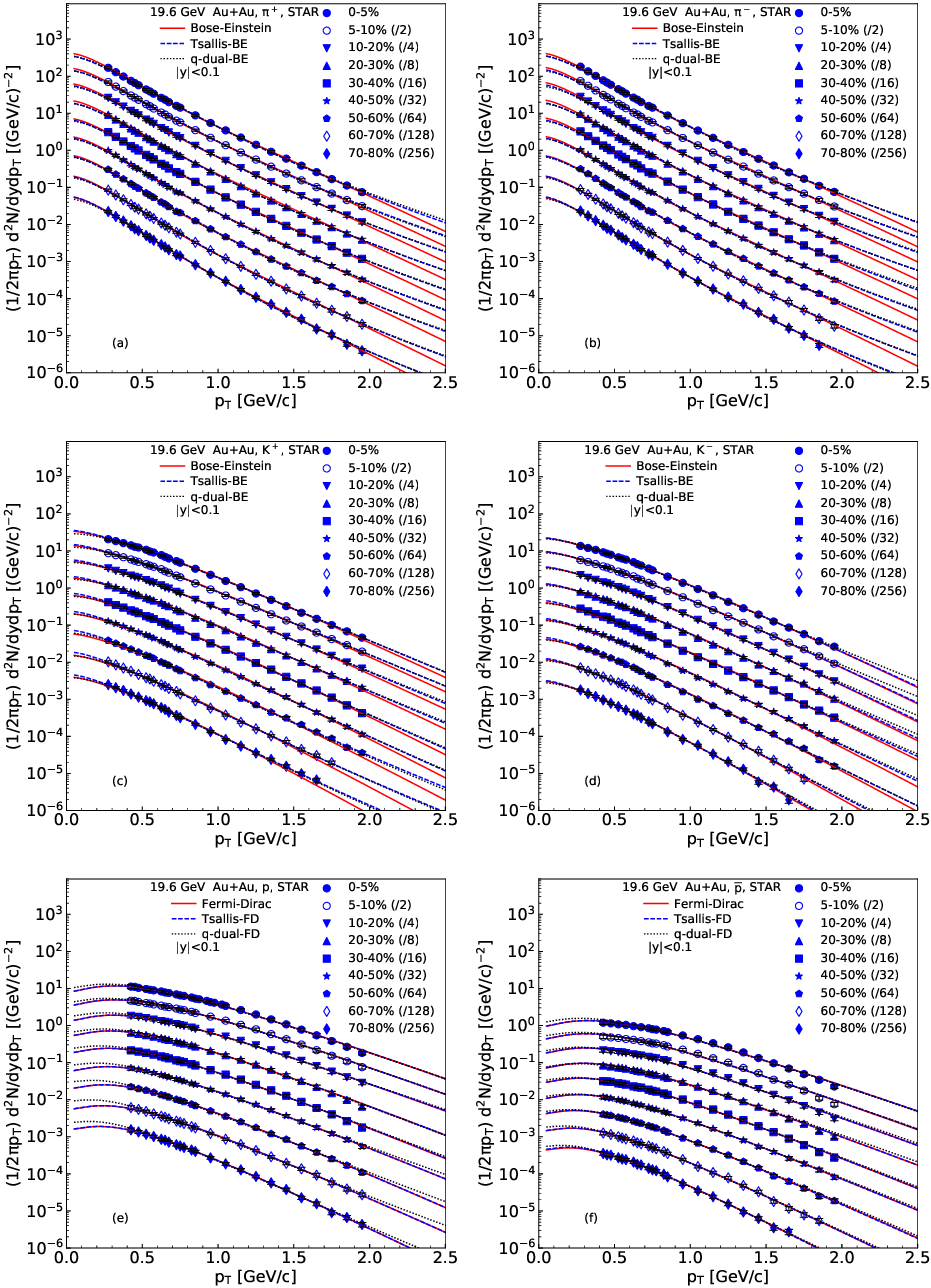}
\end{center}
\justifying\noindent {Figure 3. The invariant yields of (a)
$\pi^+$, (b) $\pi^-$, (c) $K^+$, (d) $K^-$, (e) $p$, and (f) $\bar
p$ produced in $|y|<0.1$ in Au+Au collisions at
$\sqrt{s_{NN}}=19.6$ GeV. Different symbols represent the
experimental data measured by the STAR Collaboration~\cite{23a}
with nine centrality percentage classes as those for Figure 1 and
scaled by different quantities shown in the panels. The solid,
dashed, and dotted curves are our results fitted by the
distributions from the standard, Tsallis, and q-dual statistics,
respectively.}
\end{figure*}

\begin{figure*}[htb!]
\begin{center}
\includegraphics[width=12.0cm]{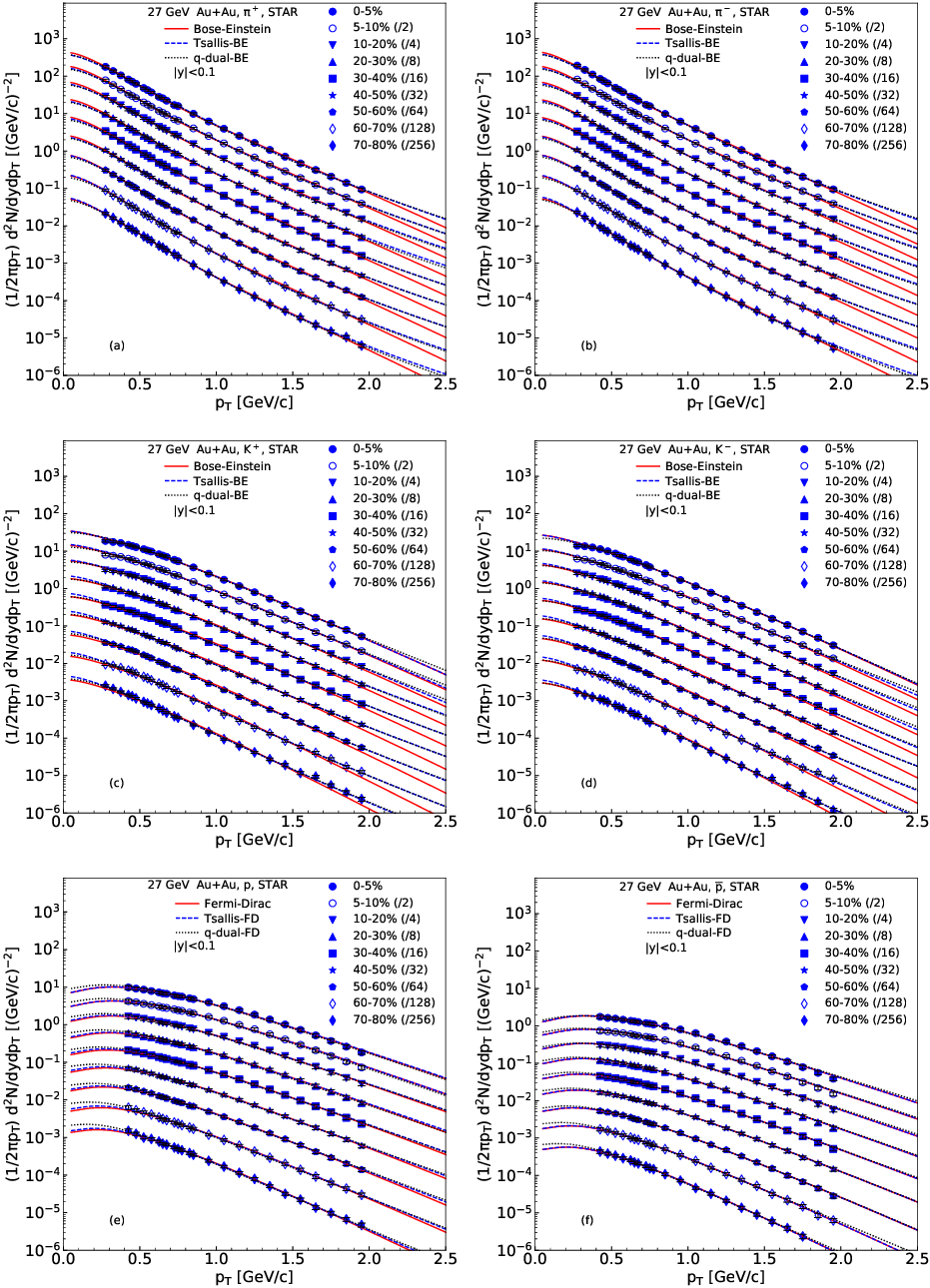}
\end{center}
\justifying\noindent {Figure 4. The invariant yields of (a)
$\pi^+$, (b) $\pi^-$, (c) $K^+$, (d) $K^-$, (e) $p$, and (f) $\bar
p$ produced in $|y|<0.1$ in Au+Au collisions at $\sqrt{s_{NN}}=27$
GeV. Different symbols represent the experimental data measured by
the STAR Collaboration~\cite{23a} with nine centrality percentage
classes as those for Figure 1 and scaled by different quantities
shown in the panels. The solid, dashed, and dotted curves are our
results fitted by the distributions from the standard, Tsallis,
and q-dual statistics, respectively.}
\end{figure*}

\begin{figure*}[htb!]
\begin{center}
\includegraphics[width=12.0cm]{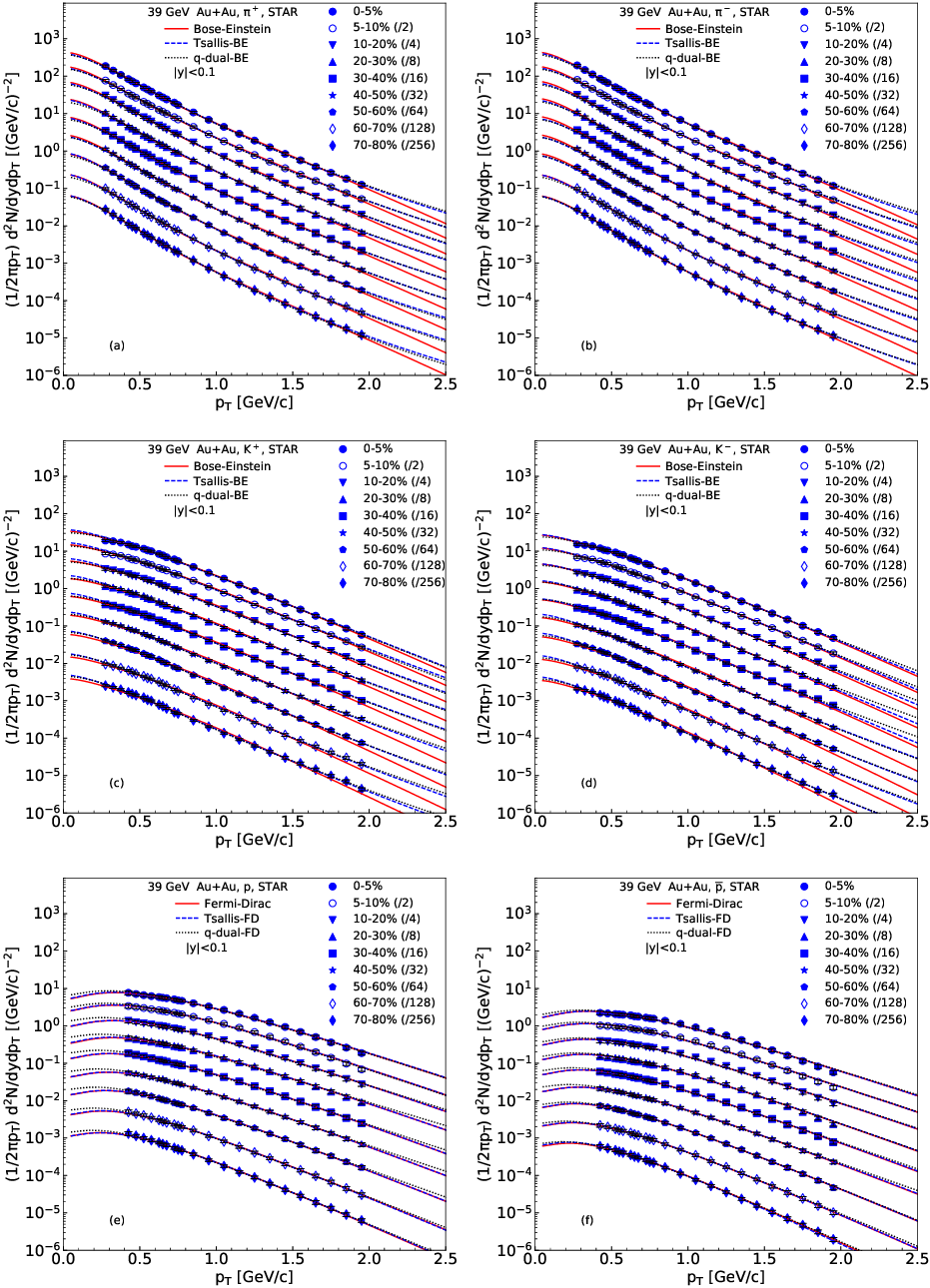}
\end{center}
\justifying\noindent {Figure 5. The invariant yields of (a)
$\pi^+$, (b) $\pi^-$, (c) $K^+$, (d) $K^-$, (e) $p$, and (f) $\bar
p$ produced in $|y|<0.1$ in Au+Au collisions at $\sqrt{s_{NN}}=39$
GeV. Different symbols represent the experimental data measured by
the STAR Collaboration~\cite{23a} with nine centrality percentage
classes as those for Figure 1 and scaled by different quantities
shown in the panels. The solid, dashed, and dotted curves are our
results fitted by the distributions from the standard, Tsallis,
and q-dual statistics, respectively.}
\end{figure*}

\begin{figure*}[htb!]
\begin{center}
\includegraphics[width=12.0cm]{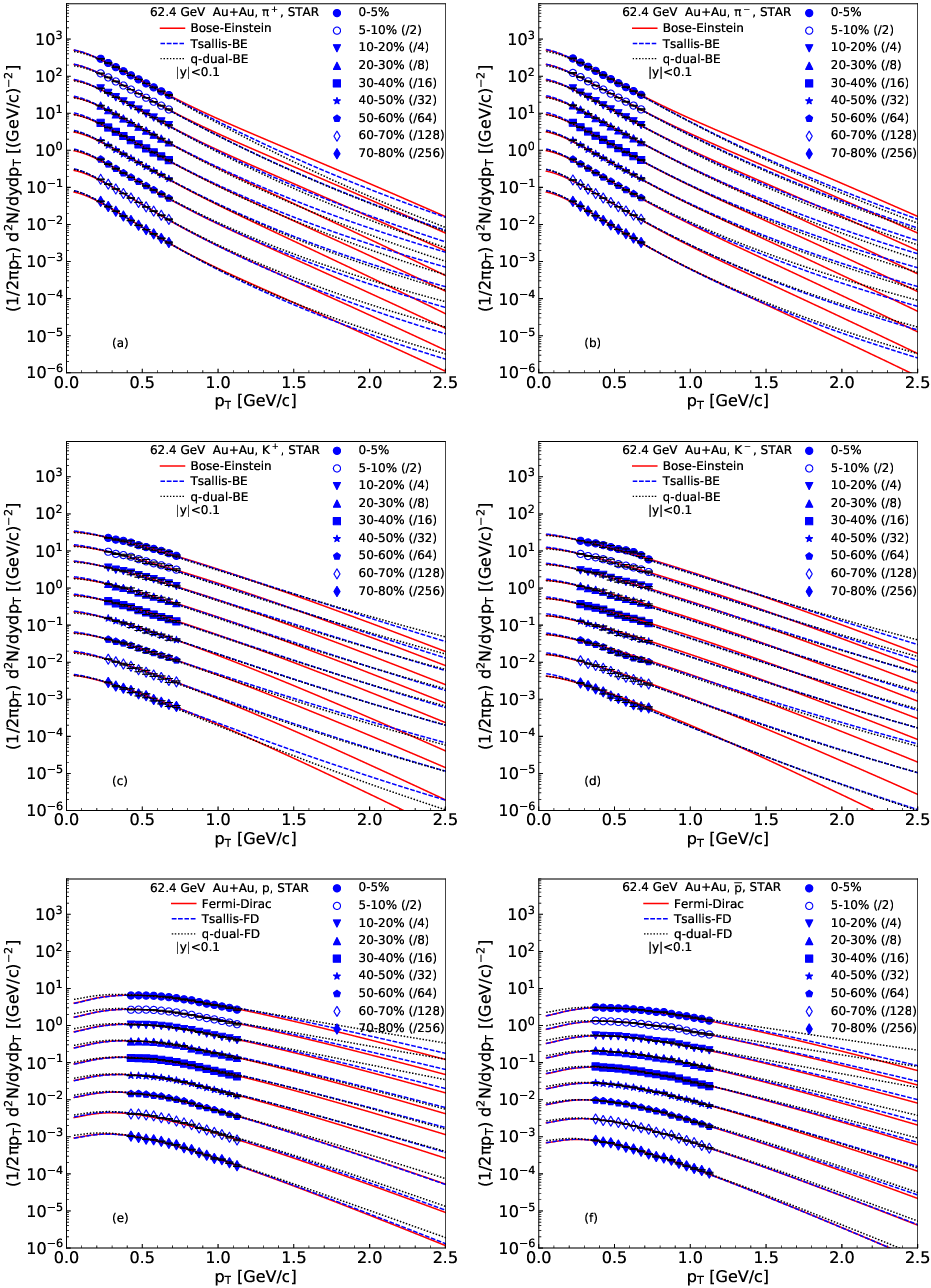}
\end{center}
\justifying\noindent {Figure 6. The invariant yields of (a)
$\pi^+$, (b) $\pi^-$, (c) $K^+$, (d) $K^-$, (e) $p$, and (f) $\bar
p$ produced in $|y|<0.1$ in Au+Au collisions at
$\sqrt{s_{NN}}=62.4$ GeV. Different symbols represent the
experimental data measured by the STAR Collaboration~\cite{23b}
with nine centrality percentage classes as those for Figure 1 and
scaled by different quantities shown in the panels. The solid,
dashed, and dotted curves are our results fitted by the
distributions from the standard, Tsallis, and q-dual statistics,
respectively.}
\end{figure*}

\begin{figure*}[htb!]
\begin{center}
\includegraphics[width=12.0cm]{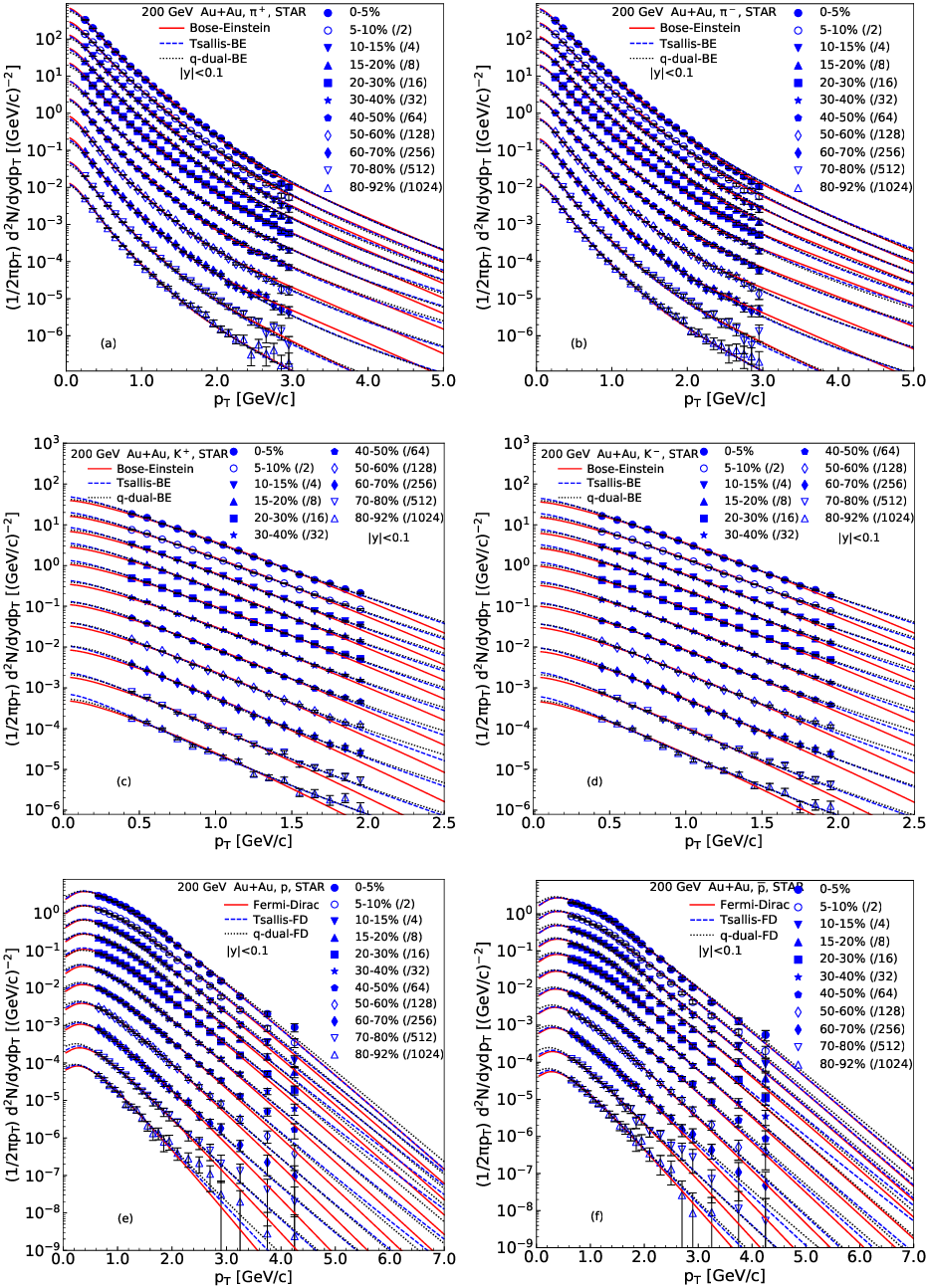}
\end{center}
\justifying\noindent {Figure 7. The invariant yields of (a)
$\pi^+$, (b) $\pi^-$, (c) $K^+$, (d) $K^-$, (e) $p$, and (f) $\bar
p$ produced in $|y|<0.1$ in Au+Au collisions at
$\sqrt{s_{NN}}=200$ GeV. Different symbols represent the
experimental data measured by the STAR Collaboration~\cite{23c}
with eleven centrality percentage classes (0--5\%, 5--10\%,
10--15\%, 15--20\%, 20--30\%, 30--40\%, 40--50\%, 50--60\%,
60--70\%, 70--80\%, and 80--92\%) and scaled by different
quantities shown in the panels. The solid, dashed, and dotted
curves are our results fitted by the distributions from the
standard, Tsallis, and q-dual statistics, respectively.}
\end{figure*}

\begin{figure*}[htb!]
\begin{center}
\includegraphics[width=12.0cm]{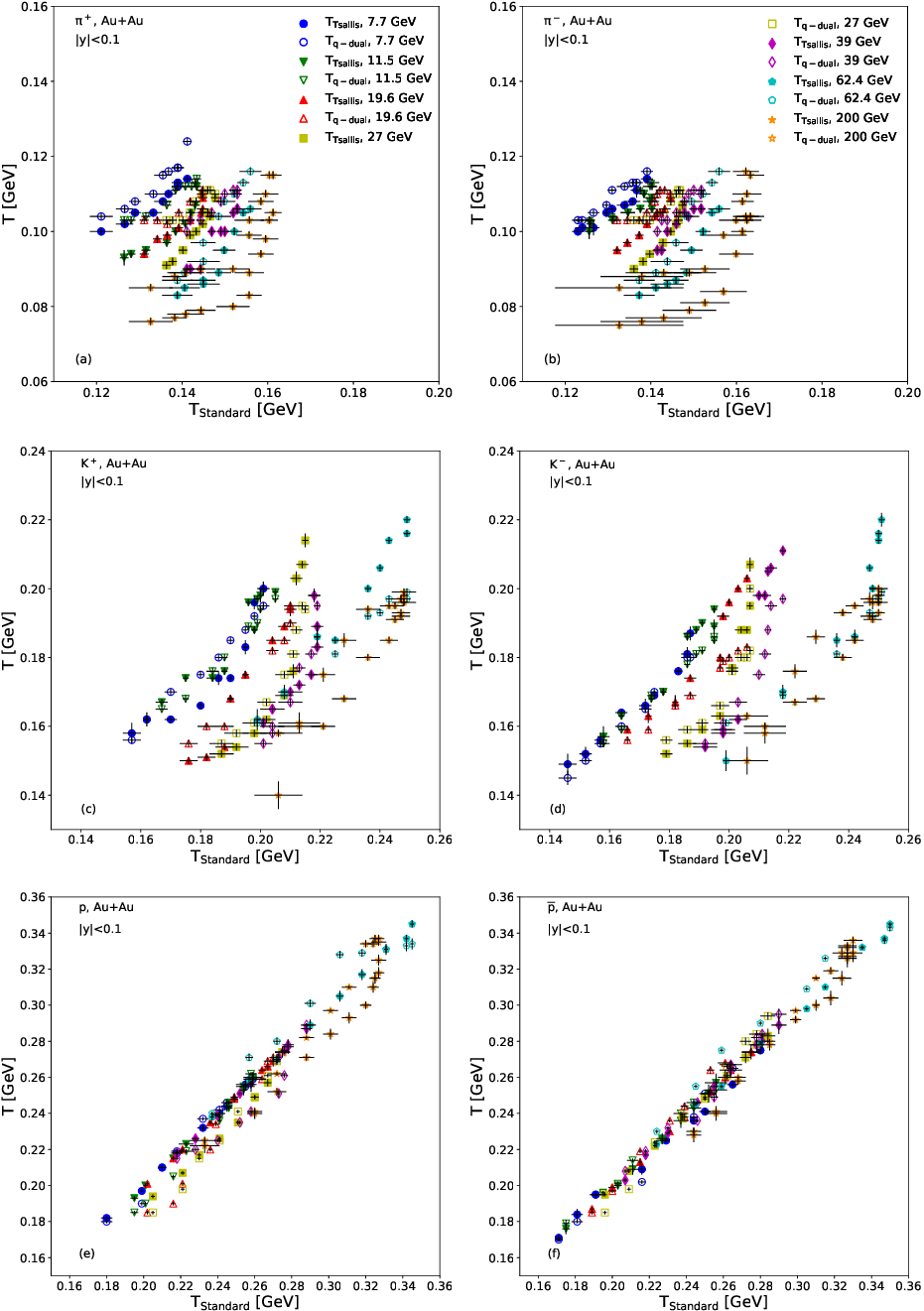}
\end{center}
\justifying\noindent {Figure 8. The dependence of
$T_{\text{Tsallis}}$ (closed symbols) and $T_{\text{q-dual}}$
(open symbols) on $T_{\text{Standard}}$ obtained from the spectra
of (a) $\pi^+$, (b) $\pi^-$, (c) $K^+$, (d) $K^-$, (e) $p$, and
(f) $\bar p$ produced in Au+Au collisions with different
centralities at RHIC. The symbols represent the results classified
by different collision energies shown in the panels.}
\end{figure*}

\begin{figure*}[htbp]
\begin{center}
\includegraphics[width=12.0cm]{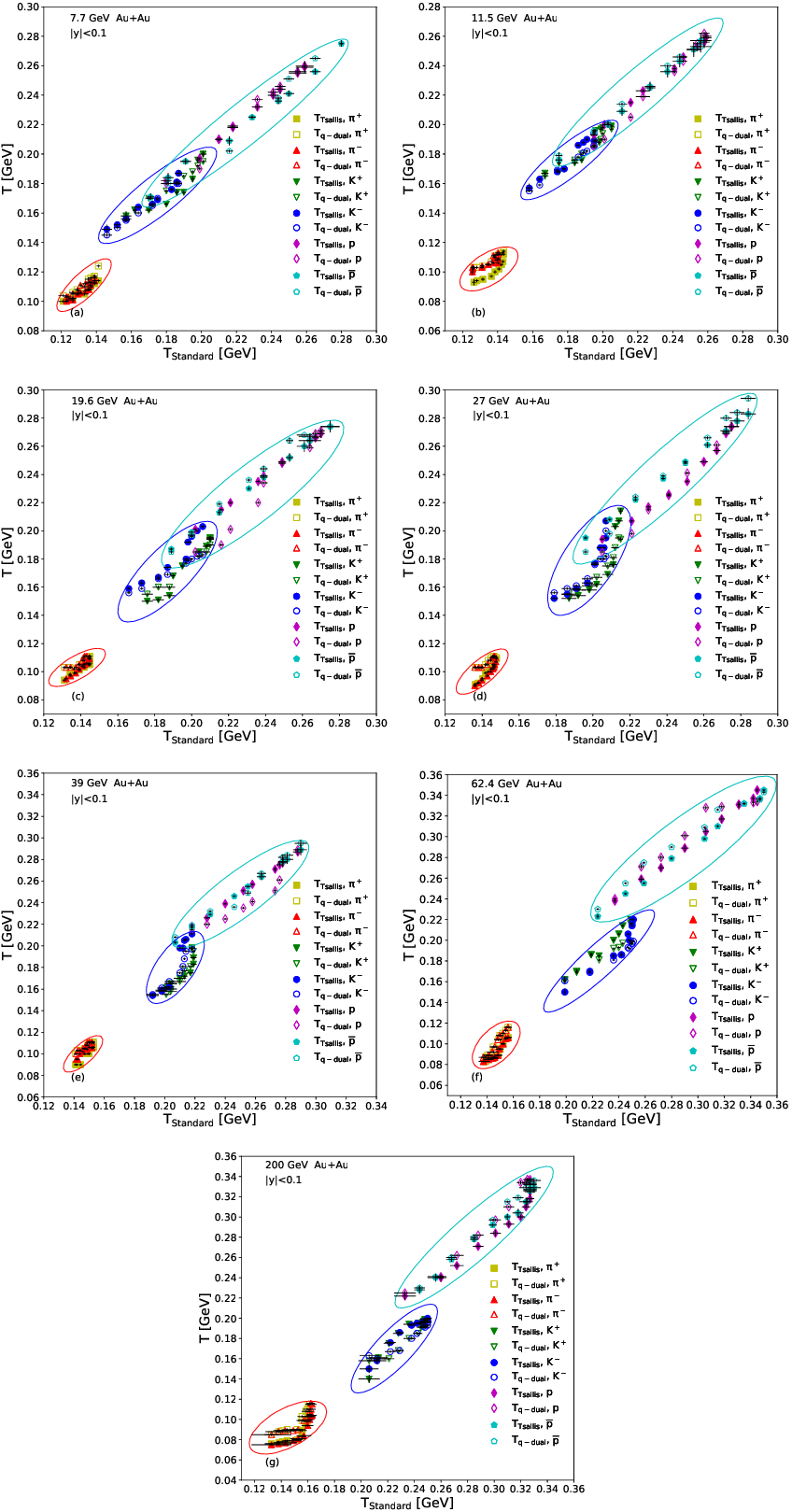}
\end{center}
\justifying\noindent {Figure 9. The dependence of
$T_{\text{Tsallis}}$ (closed symbols) and $T_{\text{q-dual}}$
(open symbols) on $T_{\text{Standard}}$ redisplayed in terms of
$\sqrt{s_{NN}}$ which equals to (a) 7.7, (b) 11.5, (c) 19.6, (d)
27, (e) 39, (f) 62.4, and (g) 200 GeV. The symbols represent the
results classified by different hadrons shown in the panels.}
\end{figure*}

Similar to the multi-component distributions in the standard
statistics, one may structure the multi-component distributions in
the Tsallis statistics which can be empirically covered by the
distributions in the q-dual statistics which has the following
related equations~\cite{40}
\begin{align}
N=gV\int \frac{d^3p}{(2\pi)^3} \sum_{k=0}^{\infty}(-S)^k
\bigg[1+\big(k+1\big)\big(q-1\big)\frac{E-\mu}{T}\bigg]^{-\frac{q}{q-1}},
\end{align}
\begin{align}
E\frac{d^3N}{d^3p} =\frac{gV}{(2\pi)^3}E \sum_{k=0}^{\infty}(-S)^k
\bigg[1+\big(k+1\big)\big(q-1\big)\frac{E-\mu}{T}\bigg]^{-\frac{q}{q-1}},
\end{align}
\begin{align}
\frac{dN}{dp}=\frac{2gV}{(2\pi)^2} p^2 \sum_{k=0}^{\infty}(-S)^k
\bigg[1+\big(k+1\big)\big(q-1\big)\frac{\sqrt{p^2+m_0^2}-\mu}{T}\bigg]^{-\frac{q}{q-1}},
\end{align}
\begin{align}
\frac{d^2N}{dydp_T} =\frac{gV}{(2\pi)^2}p_T
\sqrt{p_T^2+m_0^2}\cosh y \sum_{k=0}^{\infty}(-S)^k
\bigg[1+\big(k+1\big)\big(q-1\big)\frac{\sqrt{p_T^2+m_0^2}\cosh
y-\mu}{T} \bigg]^{-\frac{q}{q-1}},
\end{align}
\begin{align}
\frac{dN}{dp_T} =\frac{gV}{(2\pi)^2} p_T \sqrt{p_T^2+m_0^2}
\int_{y_{\min}}^{y_{\max}} \cosh y \sum_{k=0}^{\infty}(-S)^k
\bigg[1+\big(k+1\big)\big(q-1\big) \frac{\sqrt{p_T^2+m_0^2}\cosh
y-\mu}{T} \bigg]^{-\frac{q}{q-1}}dy,
\end{align}
\begin{align}
\frac{dN}{dy} =\frac{gV}{(2\pi)^2}\int_{0}^{p_{T\max}} p_T
\sqrt{p_T^2+m_0^2} \cosh y \sum_{k=0}^{\infty}(-S)^k
\bigg[1+\big(k+1\big)\big(q-1\big) \frac{\sqrt{p_T^2+m_0^2}\cosh
y-\mu}{T} \bigg]^{-\frac{q}{q-1}}dp_T,
\end{align}
where $k$ is an integer. Although the maximum $k$ is $\infty$, it
is enough if the maximum $k$ is taken to be 10 in the calculation,
where the contribution with $k>10$ is a small quantity which can
be neglected.

The experimental data analyzed herein are presented in terms of
invariant yield allowing direct comparison with modelling results.
Given that this work aims at comparing effective temperatures
derived from standard statistical mechanics alongside Tsallis and
q-dual statistics, our primary focus lies on extracting
temperature values from these three frameworks. To differentiate
between various temperatures associated with each statistical
approach effectively we will denote them as $T_{\text{Standard}}$,
$T_{\text{Tsallis}}$, and $T_{\text{q-dual}}$, orderly.
Furthermore if necessary entropy indices pertaining to Tsallis and
q-dual statistics will be referred to as $q_{\text{Tsallis}}$ and
$q_{\text{q-dual}}$ respectively.

In the extraction of effective temperatures, the chemical
potential $\mu$ is not a sensitive quantity due to its relatively
small value compared to particle energy at the relevant RHIC
energies. Nevertheless, there are at least two methods available
for obtaining $\mu$. Let $\mu_B$, $\mu_S$, and $\mu_I$ denote the
chemical potentials associated with baryon number, strangeness,
and electric charge, respectively. The chemical potential $\mu$ of
a given particle can be expressed in terms of these three chemical
potentials~\cite{43,44,46,47,48,49}. Alternatively, a more
convenient method involves using the yield ratio of negatively
charged to positively charged hadrons to determine
$\mu$~\cite{23c,50,51,53,54}. We prefer to employ this second
method as it is easier for our purposes.

\section{Results and discussion}

\subsection{Comparison with experimental data}

Figures 1--7 present the invariant yields, $(1/2\pi
p_T)d^2N/dydp_T$, for (a) $\pi^+$, (b) $\pi^-$, (c) $K^+$, (d)
$K^-$, (e) $p$, and (f) $\bar p$, produced within $|y|<0.1$ in
Au+Au collisions at $\sqrt{s_{NN}}=7.7$, 11.5, 19.6, 27, 39, 62.4,
and 200 GeV, respectively. Different symbols represent
experimental data measured by the STAR
Collaboration~\cite{23a,23b,23c} at RHIC. In Figures 1--6 we
utilize nine centrality percentage classes: 0--5\%, 5--10\%,
10--20\%, 20--30\%, 30--40\%, 40--50\%, 50--60\%, 60--70\%, and
70--80\%. Figure 7 employs eleven centrality percentage classes:
0--5\%, 5--10\%, 10--15\%, 15--20\%, 20--30\%, 30--40\%, 40--50\%,
50--60\%, 60--70\%, 70--80\%, and 80--92\%. For each centrality
class presented in these figures, the data have been scaled by
specific factors indicated in their respective panels for clarity.

In Figures 1--7, the solid, dashed, and dotted curves represent
our results fitted using distributions from standard statistics,
Tsallis statistics, and q-dual statistics, respectively. In
standard statistics, either a two-component or three-component
distribution is employed if a single-component distribution does
not adequately fit the $p_T$ spectra. Conversely, in both Tsallis
and q-dual statistics, a single-component distribution suffices.
In the fit, the least square method is performed to optimize the
parameter values. In the three statistics, the main parameters are
$T_{\text{Standard}}$, $T_{\text{Tsallis}}$, $q_{\text{Tsallis}}$,
$T_{\text{q-dual}}$, and $q_{\text{q-dual}}$. Here,
$T_{\text{Standard}}=k_1T_1+(1-k_1)T_2$ or
$T_{\text{Standard}}=k_1T_1+k_2T_2+(1-k_1-k_2)T_3$, in the case of
considering two- or three-component distribution in the standard
statistics.

In the distributions derived from both Tsallis and q-dual
statistics, the free parameters include the effective temperature
and entropy index. In contrast, for single-component distributions
based on standard statistics, only the effective temperature
serves as a free parameter. However, when considering
two-component distributions, an additional two free parameters
(temperature and fraction) are introduced; similarly,
three-component distributions result in four extra free parameters
(two temperatures and two fractions). As illustrated in Figures
1--7, experimental data from Au+Au collisions at varying
centralities and energies, measured by the STAR Collaboration at
RHIC, can be effectively fitted using distributions from standard
statistics as well as Tsallis and q-dual statistics. Notably, in
certain cases involving standard statistics, a two- or
three-component distribution is necessary.

Numerous studies have examined how collision energy, particle
mass, and event centrality influence effective temperature,
kinetic freeze-out temperature, and entropy index; these findings
have been documented in our previous work~\cite{56,57,58} as well
as other literature~\cite{59,60,61}. It is worth noting that
applications of q-dual statistics remain relatively scarce. The
aforementioned studies indicate that effective temperatures and
kinetic freeze-out temperatures exhibit slight increases or
decreases, or may even saturate, as collision energy rises beyond
7.7 GeV. Additionally, these temperatures tend to increase with
greater particle mass and event centrality levels; conversely, the
entropy index shows an upward trend with increasing collision
energy while decreasing with higher particle mass and event
centrality.

This study does not present results pertaining to any specific
exceptions previously discussed elsewhere; instead it focuses on
comparing effective temperatures obtained through three types of
statistical frameworks, specifically highlighting our limited use
of q-dual statistics. While we have previously compared effective
temperatures derived from Maxwell-Boltzmann statistics with those
from Tsallis statistics~\cite{62,63}, this work aims to explore
relationships among more extensive sets of effective temperatures
and select a baseline from these comparisons. Naturally, the
standard statistics can be regarded as the baseline due to its
solid foundation and wide applicability.

\subsection{Tendencies of parameters}

The dependence of $T_{\text{Tsallis}}$ (closed symbols) and
$T_{\text{q-dual}}$ (open symbols) on $T_{\text{Standard}}$,
derived from the spectra of (a) $\pi^+$, (b) $\pi^-$, (c) $K^+$,
(d) $K^-$, (e) $p$, and (f) $\bar p$ produced in Au+Au collisions
with varying centralities at RHIC, is illustrated in Figure 8. In
this figure, the symbols represent results categorized by
different collision energies as shown in the panels. An
alternative representation of these temperatures is provided in
Figure 9, where the dependence of both $T_{\text{Tsallis}}$ and
$T_{\text{q-dual}}$ on $T_{\text{Standard}}$ is displayed as a
function of $\sqrt{s_{NN}}$ that equals to (a) 7.7, (b) 11.5, (c)
19.6, (d) 27, (e) 39, (f) 62.4, and (g) 200 GeV. Here again, the
symbols denote results classified according to various hadrons
depicted in the panels. The oblique elliptic curves serve to guide
the eyes and differentiate between results for distinct particles.

From Figures 8 and 9, it can be observed that during the
production of $\pi^{\pm}$ in Au+Au collisions at energies of
$\sqrt{s_{NN}}=62.4$ and 200 GeV, there is a notable increase in
$T_{\text{Standard}}$, from peripheral collisions characterized by
centrality ranges of 70--80\% or 80--92\% to semi-central
collisions possed centrality of 40--50\%; however it subsequently
changes slowly when transitioning further to central collisions
carried centrality of 0--5\%. Such behavior may imply potential
formation conditions for quark-gluon plasma (QGP), analogous to
processes like ice melting or water boiling.

In contrast to this trend for $T_{\text{Standard}}$, both
$T_{\text{Tsallis}}$ and $T_{\text{q-dual}}$ exhibit significant
increases at these specific energies and centralities examined
here. Furthermore, a positive correlation appears evident between
$T_{\text{Tsallis}}$ and $T_{\text {Standard}}$, as well as
between $T _{\text{q-dual}}$ and $T_{\text{Standard}}$. This
relationship holds true across other causes at different collision
energies. The linear approximation may be disrupted due to entropy
index influences inherent within both frameworks of Tsallis and
q-dual statistics, regardless of whether QGP is formed or not.

In the production of $\pi^{\pm}$ in Au+Au collisions with varying
centralities at $\sqrt{s_{NN}}=7.7$--39 GeV, as well as in the
production of $K^{\pm}$ at $\sqrt{s_{NN}}=7.7$--200 GeV,
approximate linear relationships between $T_{\text{Tsallis}}$ and
$T_{\text{Standard}}$, along with $T_{\text{q-dual}}$ and
$T_{\text{Standard}}$, exhibit dispersion across different
collision energies. Conversely, for the production of $p(\bar p)$
in Au+Au collisions with differing centralities at
$\sqrt{s_{NN}}=7.7$--200 GeV, these relationships appear more
compact or nearly coincident across various collision energies.
This discrepancy underscores the complexity inherent to
high-energy collisions; however, it is noteworthy that the average
temperature weighted by different particles closely aligns with
that derived from the spectra of $\pi^{\pm}$ due to their
significantly high yield within this energy range.

Given that both $T_{\text{Tsallis}}$ and $T_{\text{q-dual}}$
demonstrate a significant growth when increasing centrality at
62.4 and 200 GeV, they are inadequate for accurately describing
QGP formation in central and semi-central collisions at elevated
energies. In contrast, $T_{\text{Standard}}$ exhibits a rapid
increase from peripheral to semi-central collisions before
becoming slowly increase from semi-central to central collisions.
This behavior of changing the slope relative to centrality is more
suitable for characterizing QGP formation; nonetheless, it
necessitates consideration of two- or three-component
distributions. We contend that standard statistics, being
fundamental, should receive greater emphasis in future research
endeavors involving multi-component distributions.

\subsection{Further discussions}

It is important to note that the experimental data utilized in
this analysis encompass only a limited range of $p_T$, which seems
to restrict the accurate determination of certain parameters.
However, it is understood that high-$p_T$ particles are
predominantly produced through hard scattering process, rather
than soft excitation process. Generally, hard process should not
contribute to the thermal parameters. If the experimental data
were to cover a broader range of $p_T$, one might consider
excluding high-$p_T$ particles in order to obtain thermal
parameters as accurately as possible.

Furthermore, when compared with low-$p_T$ particle yield, the
yield of high-$p_T$ particles is much smaller and can often be
disregarded. Even if the contribution of high-$p_T$ are not
excluded in extracting thermal parameters, their effects can be
neglected. Indeed, the hard scattering is associated with the
heavy tail observed in the data. While this observation holds true
for distributions derived from standard statistics, it is
important to note that distributions based on Tsallis and q-dual
statistics can effectively describe the nearly entire range of
$p_T$ distribution. Nevertheless, as we have indicated, both
including and excluding high $p_T$ spectra have minimal impact on
the extraction of thermal parameters.

Usually, the peak of the $p_T$ distribution occurs in very low
region, in which the yield is primarily influenced by strong
decays from high-mass resonances and weak decays from heavy flavor
hadrons~\cite{64}, both of which should be excluded from thermal
distributions. Conversely, effective temperature is mainly
determined by the inverse slope of the distribution arising from
soft excitation process. The absence of data covering this peak
does not significantly impact our ability to determine temperature
accurately across all methodologies employed in this study. In
fact, similar approaches have been documented in previous
literature such as refs.~\cite{23a,23b,23c}.

As discussed in the preceding section, the entropy index plays a
crucial role in extracting effective temperature within Tsallis
and q-dual statistics frameworks. Several studies have highlighted
correlations among adjustable parameters~\cite{65,66,67},
particularly within Tsallis statistics and similar non-extensive
statistical models; these studies indicate a negative correlation
between effective temperature and entropy index. These
correlations influence the temperature values obtained, which are
a key parameter in this study. To mitigate the impact of
correlations among different parameters, one not only applies the
least squares method but also allows for varying selected $p_T$
ranges for different particles when extracting thermal
parameters~\cite{23a,23b,23c}. What we do in this study is the
application of the least squares method.

While a standard distribution can effectively describe a single
source at a given excitation degree or temperature, a
multi-component standard distribution characterizes the system
with multiple sources or temperatures that reflect temperature
fluctuation occurring among these sources~\cite{68}. Due to
differing temperatures, interactions between various sources may
occur through heat energy exchange. Consequently, the collision
system evolves toward an equilibrium state; however, it ultimately
remains in an approximate equilibrium state. The exchange of heat
energy among distinct sources leads to coupling of their entropy
functions. Thus, the total entropy is represented as the sum of
entropies from various sources along with their couplings.

The multi-component standard distribution can be described using
non-extensive distributions derived from Tsallis statistics. In
explaining the origin of this non-extensive distribution,
temperature fluctuation within the multi-component standard
distribution is a candidate~\cite{68}, which corresponds to
specific temperature in Tsallis statistics. The degree of
non-equilibrium among different sources is characterized by an
entropy index within this framework. Given that multiple sources
are produced in high-energy collisions, it follows that
non-extensive Tsallis statistics may represent one of the
fundamental features inherent in such process~\cite{65}.
Furthermore, there exists a relationship between non-extensive
Tsallis distributions and Boltzmann's factor through continuous
summation weighted by factors defined by Euler--Gamma
function~\cite{68}.

Before presenting our summary and conclusions, we would like to
emphasize that the introduction of multi-source mechanisms within
the multi-source thermal model~\cite{24,25,26} does not
significantly alter the relationships among $T_{\text{Tsallis}}$,
$T_{\text{q-dual}}$, and $T_{\text{Standard}}$. In our view, these
temperature relationships are inherent due to the specific forms
used in modelling $p_T$ distributions. The incorporation of
multi-source mechanisms may yield a multi-component distribution
capable of fitting experimental $p_T$ spectra with high accuracy.
Furthermore, introducing these mechanisms can provide a clearer
physical picture that reflects the complexities involved in
high-energy collisions.

\section{Summary and conclusions}

The invariant yields of $\pi^{\pm}$, $K^{\pm}$, and $p(\bar p)$
produced in $|y|<0.1$ in Au+Au collisions with various
centralities at $\sqrt{s_{NN}}=7.7$, 11.5, 19.6, 27, 39, 62.4, and
200 GeV measured by the STAR Collaboration at RHIC have been
analyzed using distributions derived from standard statistics as
well as Tsallis and q-dual statistics. In certain cases, a two- or
three-component distribution within standard statistics is
necessary for accurate representation. The experimental data can
be effectively fitted using these related distributions. Effective
temperature parameters such as $T_{\text{Standard}}$,
$T_{\text{Tsallis}}$, and $T_{\text{q-dual}}$ are extracted
through fitting the spectra of identified light charged hadrons.

In the production of $\pi^{\pm}$ in collisions at
$\sqrt{s_{NN}}=62.4$ and 200 GeV, it is observed that while
$T_{\text{Standard}}$ increases rapidly initially before changing
more gradually, both $T_{\text{Tsallis}}$ and $T_{\text{q-dual}}$
exhibit significant increases with rising centrality levels. The
anticipated approximate linear relationships between
$T_{\text{Tsallis}}$ and $T_{\text{Standard}}$, as well as between
$T_{\text{q-dual}}$ and $T_{\text{Standard}}$, are disrupted due
to the influence of entropy index associated with QGP formation;
however, similar approximate linear relationships do exist in
other scenarios.

For the production of $\pi^{\pm}$ across different centralities at
$\sqrt{s_{NN}}=7.7$--39 GeV, along with that of ${K}^{\pm}$ over a
range from $\sqrt{s_{NN}}=7.7$ to 200 GeV, there appears to be a
dispersion in the approximate linear relationships between both
$T_\text{Tsallis}$ versus $T_\text{Standard}$, as well as
$T_\text{q-dual}$ versus $T_\text{Standard}$ across varying
collision energies. Conversely, for $p(\bar p)$ production under
different centralities spanning from $\sqrt{s_{NN}}=7.7$ to 200
GeV, these aforementioned approximate linear relationships remain
compacted or nearly coincide across distinct collision energies.

If QGP is formed in central and semi-central Au+Au collisions at
$\sqrt{s_{NN}}=62.4$ and 200 GeV, one would expect the temperature
to display a slowly changing trend as a function of centrality.
However, both $T_{\text{Tsallis}}$ and $T_{\text{q-dual}}$ exhibit
a significant increase, and $T_{\text{Standard}}$ demonstrates a
slowly increase, with increasing centrality. We argue that
standard statistics, characterized by multi-component
distributions, are more appropriate for describing QGP formation
and should receive greater emphasis in future research.
\\
\\
{\bf Acknowledgements}

The work of Shanxi Group was supported by the National Natural
Science Foundation of China under Grant No. 12147215, the Shanxi
Provincial Basic Research Program (Natural Science Foundation)
under Grant No. 202103021224036, and the Fund for Shanxi ``1331
Project" Key Subjects Construction. The work of P.P.Y. was
supported by the Shanxi Provincial Basic Research Program (Natural
Science Foundation) under Grant No. 202203021222308, the Doctoral
Scientific Research Foundations of Shanxi Province and Xinzhou
Normal University, and the Academic Leading Specialist Project of
Xinzhou Normal University under Grant Nos. 2024RC10 and 2024RC10B.
The work of K.K.O. was supported by the Agency of Innovative
Development under the Ministry of Higher Education, Science and
Innovations of the Republic of Uzbekistan within the fundamental
project No. F3-20200929146 on analysis of open data on heavy-ion
collisions at RHIC and LHC.
\\
\\
{\bf Data Availability Statement}\hskip0.35cm This manuscript has
associated data in a data repository. [Author's comment: The data
analyzed in this manuscript were obtained from
https://www.hepdata.net/, a freely accessible repository.]
\\
\\
{\bf Code Availability Statement}\hskip0.35cm This manuscript has
no associated code/software. [Author's comment: Code/Software
sharing not applicable to this article as no code/software was
generated or analysed during the current study].
\\
\\
{\bf Disclosure}\hskip0.35cm The funding agencies have no role in
the design of the study; in the collection, analysis, or
interpretation of the data; in the writing of the manuscript; or
in the decision to publish the results.
\\
\\
{\bf Conflicts of Interest}\hskip0.35cm The authors declare that
there are no conflict of interest regarding the publication of
this paper.
\\
\\
{\bf Ethical Standards}\hskip0.35cm This research complies with
the ethical standards.

\end{document}